\documentclass[a4paper,12pt]{article}
\pdfoutput=1
\usepackage[dvipdfmx]{graphicx}
\usepackage{cite}
\usepackage{amsmath}
\usepackage{amsfonts}
\usepackage{amssymb}
\usepackage{graphicx, rotating}
\usepackage{epsfig}
\usepackage{latexsym}
\usepackage{graphicx}
\usepackage{color}
\usepackage{amsmath,bm,amssymb}
\usepackage{slashed}
\usepackage{color}
\usepackage[english]{babel}
\usepackage{hyperref}

\newcommand{\Slash}[1]{{\ooalign{\hfil#1\hfil\crcr\raise.167ex\hbox{/}}}}

\newcommand{\beq}{\begin{equation}}  \newcommand{\eeq}{\end{equation}}
\newcommand{\bef}{\begin{figure}}  \newcommand{\eef}{\end{figure}}
\newcommand{\bec}{\begin{center}}  \newcommand{\eec}{\end{center}}
  
\newcommand{\laq}[1]{\label{eq:#1}}  

\newcommand{\Eq}[1]{Eq.~(\ref{eq:#1})}

\newcommand{\eq}[1]{(\ref{eq:#1})}

\newcommand{\ab}[1]{\left|{#1}\right|}
\newcommand{\vev}[1]{ \left\langle {#1} \right\rangle }

\newcommand{\SU}[1]{{\rm SU{#1} } }

\def\({\left(}
\def\){\right)}

\def\O{\mathcal{O}}
\def\U{\mathop{\rm U}}

\def\tr{\mathop{\rm tr}}

\newcommand{\AND}{~{\rm and}~}

\newcommand{\GEV}{ {\rm \, GeV} }

\def\a{\alpha}

\def\d{\delta}
\def\e{\epsilon}
\def\f{\phi}

\def\m{\mu}
\def\n{\nu}

\def\x{\xi}

\def\L{\Lambda}
\def\F{\Phi}

\def\tl{\tilde}
\def\*{\dagger}

\begin{document}
\renewcommand\bibname{\Large References}

\begin{flushright}
\end{flushright}

\begin{center}

\vspace{1.5cm}

{\Large\bf {Gauge coupling jump and small instantons \\ 
from a large non-minimal coupling}}
\vspace{1.5cm}

{\bf Juntaro Wada}$^{a}$~~ and~~ {\bf Wen Yin}$^{b}$

\vspace{1.5cm}

{\em $a.$ Department of Physics, University of Tokyo, Bunkyo-ku, Tokyo, 113-0033, Japan} \\
{\em $b.$ Department of Physics, Tokyo Metropolitan University, Tokyo 192-0397, Japan}

\vspace{1.5cm}
\abstract{If a scalar field couples to the Ricci scalar with a large non-minimal coupling, the Standard Model coupling parameters can differ above and below an intermediate field range of the scalar due to the non-renormalizability. In this paper, we study, for the first time, the threshold effects on a gauge coupling in both Metric and Palatini formulations of gravity. We find that the gauge coupling naturally jumps around this intermediate scale since counter terms for the renormalization behave so. If the gauge coupling becomes strong with a large scalar field value due to this effect, there can be an enhanced small instanton contribution, the dilute gas approximation of which is justified because the gauge sector decouples when the scalar wave mode is very short.
Using these findings, we discuss the QCD axion quality problem, the heavy QCD axion, the axion abundance, and the suppression of isocurvature perturbations. 
We show that axion physics may differ substantially from na\"{i}ve expectations when we introduce a large non-minimal coupling for any scalar field.
}
\end{center}
\clearpage
\section{Introduction}
If there is a CP-even scalar field, $\phi$, in the theory, a non-minimal coupling to gravity is naturally expected, 
\begin{equation}
\laq{nonmini}
{{\cal L}\supset \frac{1}{2}M_{\rm pl}^2\Omega^2g_J^{\mu\nu}{\cal R}_{J\mu\nu},}
\end{equation}
where $g_J\AND {\cal R}_J$ are, respectively, the metric and the Ricci curvature tensor in the Jordan frame, and
\begin{equation}
\laq{Omega} 
\Omega\equiv \(1+\x \frac{ \phi^2}{M_{\rm pl}^2}\)^{1/2},
\end{equation}
with $M_{\rm{pl}} = 2.4 \times 10^{18}~\rm{GeV}$ being the reduced Planck mass. 
This simple setup has been mostly considered in the context of
cosmic inflation~\cite{Starobinsky:1979ty, Starobinsky:1980te, Guth:1980zm, Sato:1980yn, Kazanas:1980tx, Linde:1981mu, Mukhanov:1981xt, Hawking:1981fz, Chibisov:1982nx, Hawking:1982cz, Guth:1982ec, Albrecht:1982wi, Starobinsky:1982ee}, which generates primordial density perturbations, as strongly suggested by recent cosmic microwave background (CMB) data~\cite{Planck:2018jri, Planck:2018vyg}, because at very large $\phi$ field values, $M_{\rm pl}$ can be neglected and the action becomes globally scale-invariant. Thus, the potential for slow-roll inflation is flat.

The Higgs inflation~\cite{Bezrukov:2007ep, Bezrukov:2008ej} (see also \cite{Rubio:2018ogq}), in which $\phi$ is the Higgs boson, was considered as a minimal possibility for inflation. A non-minimal coupling between the Higgs and Ricci scalar, $\xi$, is much larger than unity in this scenario. In this framework, the Standard Model (SM) Yukawa and Higgs quartic couplings can differ between the current universe and the inflationary period due to so-called threshold effects~\cite{Bezrukov:2014ipa, George:2015nza, Fumagalli:2016lls, Enckell:2016xse, Bezrukov:2017dyv,Shaposhnikov:2020fdv,Poisson:2023tja}. This effect can be understood as arising from the UV-completion likely present in the intermediate region $M_{\rm pl}/\xi < \phi < M_{\rm pl}/\sqrt{\xi}$, where unitarity is violated~\cite{Burgess:2009ea, Barbon:2009ya, Burgess:2010zq, Bezrukov:2010jz}. While this primarily applies to the metric formulation, the Palatini formulation~\cite{Einstein:1925, Ferraris:1982} has a weaker perturbativity bound~\cite{Bauer:2010jg}. Still threshold effects are expected according to renormalization of the model (even if the non-minimal coupling is not large the threshold effect may exist~\cite{Poisson:2023tja}). Due to this coupling change, the electroweak (EW) vacuum has been discussed as being absolutely stabilized, making Higgs inflation possible~\cite{Bezrukov:2014ipa}. Apart from this coupling shift, the Standard Model parameters had been thought to disfavor minimal Higgs inflation because the Higgs quartic coupling turns negative at energy scales above $10^{9-12}\,\GEV$. However, it was recently shown that the EW vacuum can be made absolutely stable, and the inflation can be successful by taking into account the would-be hilltop of the Higgs potential and a very large non-minimal coupling\cite{Yin:2022fgo}.\footnote{One can have smaller non-minimal coupling if the quartic coupling turns negative around $ 10^{16}(10^{14})\GEV$ with the metric (Palatini) formulation by using the same mechanism~\cite{Yin:2022fgo, Masina:2024ybn}.}

The non-minimal coupling can also be incorporated into the Peccei-Quinn (PQ) model~\cite{Peccei:1977ur, Peccei:1977hh}, which is a promising candidate for resolving the strong CP problem. As is well known, in this model, the problem can be addressed through the axion~\cite{Weinberg:1977ma, Wilczek:1977pj}, the pseudo Nambu-Goldstone boson that emerges from the spontaneous breaking of the $U(1)_{\mathrm{PQ}}$ symmetry. However, the PQ model is known to have the quality problem~\cite{Kamionkowski:1992mf, Holman:1992us, Barr:1992qq}, 
the notion of why the PQ symmetry is so precise to solve the strong CP problem. 
In particular, all global symmetries are known to be violated by quantum gravity effects~\cite{Hawking:1987mz, Lavrelashvili:1987jg, Giddings:1988cx, Coleman:1988tj, Gilbert:1989nq, Banks:2010zn}. 
Recently, it has been demonstrated that the axion quality problem induced by gravitational wormholes can be alleviated with the non-minimal couplings of the PQ Higgs field~\cite{Hamaguchi:2021mmt, Cheong:2022ikv, Cheong:2023hrj, Kanazawa:2023xzy}.\footnote{Non-minimal coupling to gravity is also discussed in the context of axion isocurvature problem~\cite{Folkerts:2013tua, Takahashi:2015waa, Berbig:2024ufe}.} 

Even if the gravitational one is absent, the anomalous PQ symmetry suffers from the small QCD instanton contributions, which are another source inducing the quality problem~ \cite{Dine:1986bg, Kitano:2021fdl}. 
The small instanton arises in various models, e.g. models altering the running of the gauge coupling~\cite{Holdom:1982ex, Holdom:1985vx, Flynn:1987rs}, models where QCD is embedded into either a higher-dimensional theory \cite{Poppitz:2002ac, Gherghetta:2020keg, Kitano:2021fdl} and into a largerer gauge theory~\cite{Agrawal:2017ksf, Agrawal:2017evu, Fuentes-Martin:2019bue, Csaki:2019vte, Takahashi:2021tff, Babu:2024udi}, and also models with composite dynamics~\cite{Gherghetta:2020ofz, Aoki:2024usv}. 
To resolve the quality problem, one considers the PQ symmetry as some accidiental symmetries of renormalizable models, such as the ``baryon number" in the SU(N) gauge theories~\cite{DiLuzio:2017tjx, Lee:2018yak, Ardu:2020qmo, Yin:2020dfn}.  
More recently, apart from the quality problem, the CP-violating small $\SU(2)$ weak instanton was pointed out to enhance the baryon number violation for having successful baryogenesis with a lower reheating temperature than the EW scale~\cite{Jaeckel:2022osh}.

In this paper, we point out that in the presence of a large non-minimal coupling to a scalar, there is a threshold effect, naturally causing the shift of the coupling when the scalar field value is around the intermediate scale similar to the previously mentioned behaviors of the Yukawa and Higgs quartic couplings.
We show that the small instanton contribution can be enhanced if the gauge coupling becomes strong in the UV regime due to this effect. Usually, instanton calculation with field-dependent gauge coupling is not easy to perform, but in our case, the high momentum modes of the scalar field, which is typically large in the real space, decouples from the gauge sector. 
This nature allows us to use a dilute gas approximation for the high momentum mode, and it turns out that there is a significant small instanton contribution. 
By considering the gauge field being the gluon, we apply this result to discuss the PQ quality problem, heavy QCD axion, axion abundance, and axion isocurvature problem in the Higgs inflation scenario.

The organization of this paper is as follows: In Section~\ref{sec: QCD gauge coupling jump}, we explain the threshold effect and how it induces a shift in the QCD gauge coupling in the Higgs non-minimal coupling case. We also discuss the possibility of a stronger QCD during inflation due to this threshold effect.  In Section~\ref{sec: small instanton}, we argue for the existence of small instantons due to the QCD coupling jump by employing the Wilsonian approach. Finally, Section~\ref{sec: conclusions} presents the concluding remarks and generalization to other scalars with non-minimal couplings.

\section{Gauge coupling jump from a large non-minimal coupling}
\label{sec: QCD gauge coupling jump}
In this section, we discuss the threshold effect on a gauge coupling due to a large non-minimal coupling. For concreteness, we focus on the QCD coupling and the non-minimal coupling of the Higgs field $H$, i.e., $\phi = H$ in Eq.~\eqref{eq:nonmini}. 
Our discussion can be easily extended to other gauge couplings and scalar fields, and we will address other cases at the end of Sec.~\ref{sec: conclusions}.

\subsection{Framework}

The relevant parts of the Lagrangian are given by $\cal{L}_{\rm{Higgs}} + \cal{L}_{\rm{gluon}} + \cal{L}_{\rm{quark}}$, where 
\begin{align}
\mathcal{L}_{\rm{Higgs}} = \frac{1}{2}M_{\rm pl}^2\Omega^2g^{\mu\nu}_{J}{\cal R}_{J \mu\nu} + g^{ \mu\nu}_{J}\partial_{\mu} H^{\dagger}\partial_{\nu}H - U(H^\dagger H),
\end{align}
incorporates the non-minimal coupling to gravity as outlined in Eq.~\eqref{eq:nonmini} within the Jordan frame. Here, $U(H^\dagger H) = \lambda \left( H^{\dagger} H - v_{\rm{EW}}^2/2\right)^2$, with $v_{\rm{EW}} = 246~\rm{GeV}$ being the vacuum expectation value of the Higgs field. The expressions for $\mathcal{L}_{\rm{gluon}}$ and $\mathcal{L}_{\rm{quark}}$ are given by:
\begin{align}
&\mathcal{L}_{\rm gluon}= -\frac{1}{2 g^2} \tr{G_{\mu\nu}G^{\mu\nu}},\\
&\mathcal{L}_{\rm quark}=i\bar u\slashed{D} u + i\bar d\slashed{D} d - h \bar u \cdot Y_u \cdot u - h \bar d \cdot Y_d \cdot d,
\end{align}
where $D_\m$ is the covariant derivative, $g$ is QCD gauge coupling, and $Y_{u,d}$ are the Yukawa couplings for the up- and down-type quarks, respectively.
For simplicity, we adopt the unitary gauge, where the Higgs field is parameterized as $H = (0, h)/\sqrt{2}$.

To move to the Einstein frame, a Weyl transformation is performed, $g_{J \mu\nu} \to g_{\mu\nu} = \Omega^{-2} g_{J \mu\nu}$, with $\Omega \equiv \left(1 + \xi h^2/M_{\rm{pl}}^2\right)^{1/2}$, as defined in Eq.~\eqref{eq:Omega}. After this transformation, the kinetic terms for $h$, $u$, and $d$ are modified as follows:
\beq 
{\cal L}_{\rm Higgs} \supset \frac{Z[h]}{2}\partial_\mu h\partial^\mu h, ~~{\mathcal L}_{\rm quark} \supset i \Omega^{-3} \bar u\slashed{\partial} u + i \Omega^{-3} \bar d\slashed{\partial} d,
\eeq 
where $Z[h]$ depends on the formulation of gravity:
\beq 
Z[h]=
\begin{cases} 
\Omega^{-2} + \frac{3}{2}(M_{\rm pl}\frac{d}{d h}\log(\Omega^2))^2 & \text{~~(Metric)}, \\
\Omega^{-2} & \text{~~(Palatini)},
\end{cases}
\eeq 
whereas $\mathcal{L}_{\rm{gluon}}$ remains invariant under this transformation.
The canonically normalized fields, $\varphi$ and $ \psi_{u,d}$ can be obtained by normalizing the wave function of $h,u,d$ in the Einstein frame, via the relation $d \varphi/ dh = Z[h]^{1/2}$ and $\psi_{u,d}=\Omega^{3/2} u,\Omega^{3/2} d$ respectivity. In terms of these canonically normalized fields, the relevant piece of the Lagrangian $\cal{L}_{\rm{Higgs}} + \cal{L}_{\rm{gluon}} + \cal{L}_{\rm{quark}}$ becomes
\begin{align}
&\mathcal L_{\rm Higgs} 
= \frac{1}{2}M_{\rm pl}^2 R + \frac{1}{2}(\partial \varphi)^2 - V(\varphi), \\
&\mathcal{L}_{\rm gluon}= -\frac{1}{2 g^2} \tr{G_{\mu\nu}G^{\mu\nu}},\\
&\mathcal{L}_{\rm quark}=i\bar \psi_u \slashed{D} \psi_u + i\bar \psi_d \slashed{D} \psi_d - \frac{h(\varphi)}{\Omega(\varphi)} \left(\bar{\psi}_u\cdot Y_u \cdot \psi_{u} + \bar{\psi}_d\cdot Y_d \cdot \psi_{d}\right), 
\end{align}
where $R$ denotes the Ricci scalar in the Einstein frame, $R \equiv g^{\mu\nu} R_{\mu\nu}$ with $g_{\m\n} \AND R_{\m\n}$, respectively, being the Einstein frame metric and Ricci tensor, and $V(\varphi) \equiv U(\varphi)/\Omega(\varphi)^4$ represents the Higgs potential in the Einstein frame.

\subsection{Threshold effect for the QCD gauge coupling}
\label{subsec: Threshold correction for the QCD gauge coupling}
With a large Higgs non-minimal coupling, shifts in the couplings across intermediate regime are predicted~\cite{Bezrukov:2014ipa, George:2015nza, Fumagalli:2016lls, Enckell:2016xse, Bezrukov:2017dyv,Shaposhnikov:2020fdv,Poisson:2023tja}. This phenomenon has been studied in the context of Yukawa and Higgs quartic couplings~\cite{Bezrukov:2014ipa}.
Here, we estimate the impact of quark Yukawa interactions, particularly those involving the top quark, on the QCD gauge coupling jump through loop corrections (see Refs.\,\cite{Ford:1992pn,DeSimone:2008ei,Buttazzo:2013uya} for expressions of various two-loop corrections).

As mentioned previously, the tree-level Yukawa interaction is given by 
\begin{align}
{\mathcal L}_{\rm quark} \supset &\frac{h(\varphi)}{\Omega(\varphi)} \left(\bar{\psi}_u\cdot Y_u \cdot \psi_{u} + \bar{\psi}_d\cdot Y_d \cdot \psi_{d}\right)\\
\equiv  &F \left(\bar{\psi}_u \cdot Y_u \cdot \psi_u +\bar{\psi}_d\cdot Y_d \cdot \psi_{d} \right),\laq{Fdef}
\end{align}
where $F(\varphi)$ is a function of $h$ and thus $\varphi$. It can be approximated as $F(\varphi) \simeq h$ for small field values of $h$. 
We expand the Higgs field around its background value as $\varphi = \bar{\varphi} + \delta \varphi$ and proceed with loop calculations to estimate the wave function renormalization of the QCD gluon field.

\begin{figure}[t!]
\begin{center}  
\includegraphics[width=105mm]{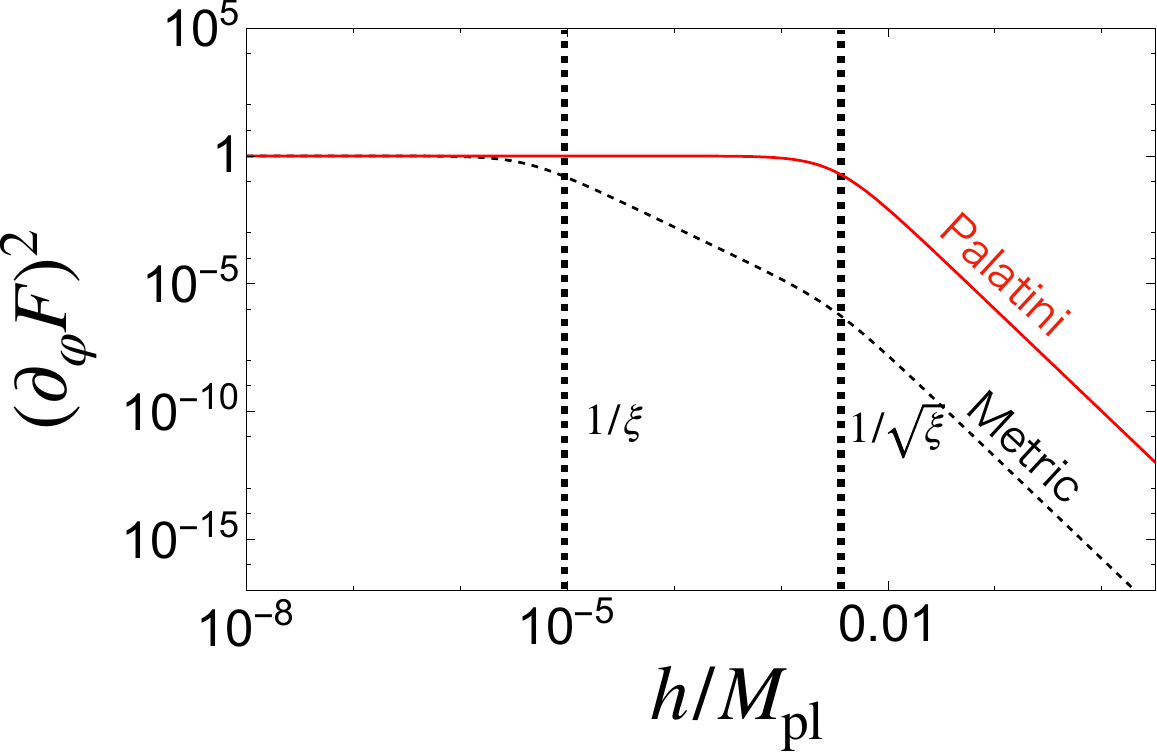}
\includegraphics[width=105mm]{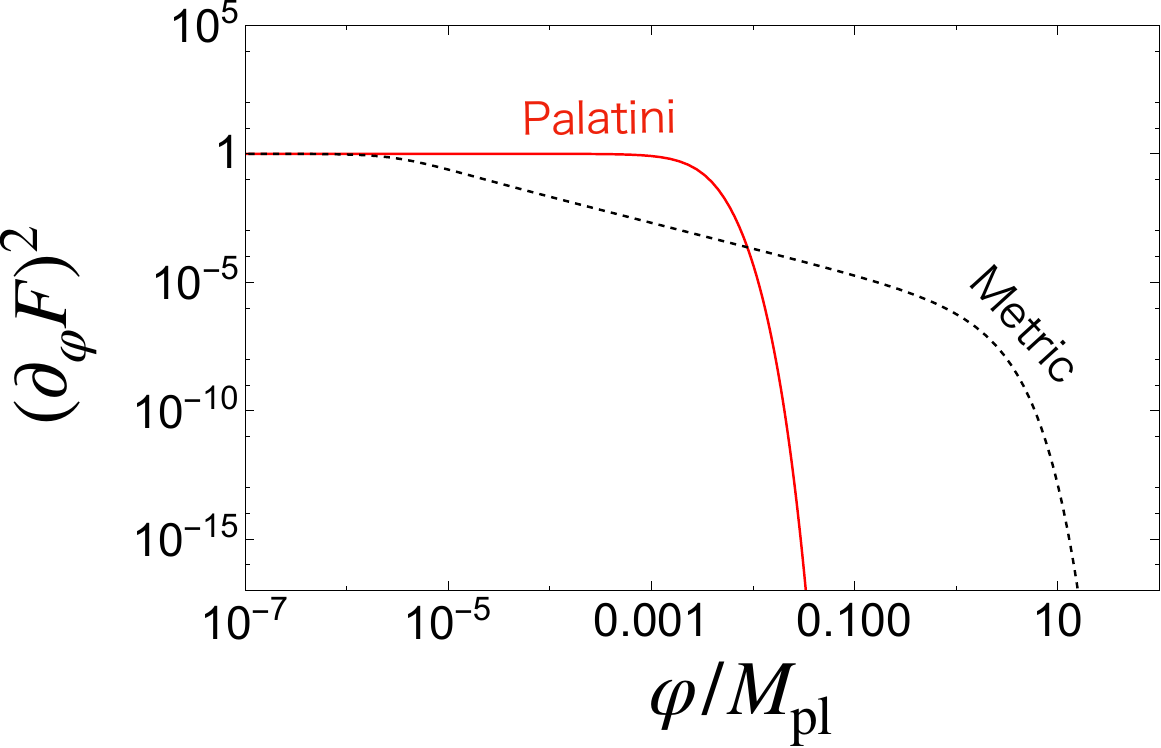}
\end{center}
\caption{$(\partial_\varphi F)^2$, relevant to the correction to the gauge coupling$^{-2}$ by varying $h/M_{\rm pl}$ (top panel) and the canonically normalized $\varphi$ (bottom panel). We take $\x=10^5$ in both figures. The black and red lines denote the case of metric and Palatini formulation respectively. 
The vertical lines in the top panel denote the $h/M_{\rm pl}=1/\x, 1/\sqrt{\x}$ from left to right. 
}
\label{fig:1}
\end{figure}

The loop-induced divergence from the Feynman diagram shown in Fig.~\ref{fig:1} is proportional to 
\beq
 \propto {\rm tr}G_{\m \n} G^{\m \n}\(-\frac{2}{(4\pi)^4} (\partial_\varphi F)^2\({\rm tr}{Y_u\cdot Y^\*_u}+{\rm tr}{Y_d \cdot Y^\*_d}\)\).
\label{eq: two loop correction for gluon coupling}
\eeq
In addition, we have terms $\propto \partial^2_{\varphi} F \partial^2_{\varphi} V$
with $\partial^2_{\varphi} V$ being the Einstein frame Higgs effective mass in dimensional regularization. By assuming a small Higgs effective mass, we do not consider this term. 

\begin{figure}[t!]
\begin{center}  
\includegraphics[width=90mm]{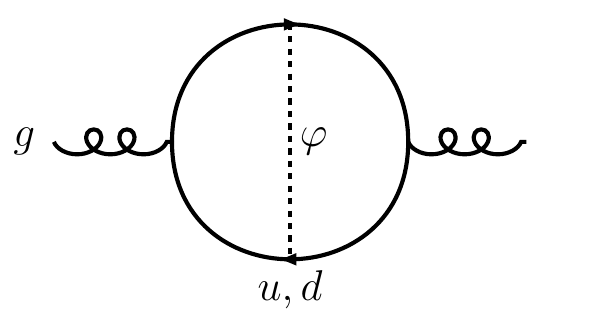}
\end{center}
\caption{Two-loop diagram for the gluon wave function. With the large non-minimal coupling, this diagram contributes a non-trivial $\varphi$-dependent divergence. 
}
\label{fig:2loop_diagram}
\end{figure}

To renormalize the divergence in Eq.~\eqref{eq: two loop correction for gluon coupling}, the following counterterms are required:
\beq
\label{counter}
\delta \mathcal{L}_{\text{ct}} \supset -\left( \partial_\varphi F \right)^2\, \mathrm{tr}\, G_{\mu \nu} G^{\mu \nu} \left( \frac{2\left( \mathrm{tr}\left[ Y_u Y_u^\dagger \right] + \mathrm{tr}\left[ Y_d Y_d^\dagger \right] \right)}{(16\pi^2)^2 (4 - d)} + \frac{1}{2}\, \delta g^{-2} \right),
\eeq
where $\delta g^{-2}$ represents an arbitrary finite constant.
We remark that this $\delta g^{-2}$ cannot be taken to zero without losing generality. In other words, we should include this $\varphi$ dependent term in the original Lagrangian similar to the higher dimensional term in the usual non-renormalizable theory.  The size of our ``higher dimensional term" defines the theory, which may or may not be UV completed. Taking this ``higher dimensional term"  to be highly suppressed is to choose one of the very large classes of models, and so it is natural to keep it.  

In Fig.~\ref{fig:1}, we plot $(\partial_\varphi F)^2$ which is the contribution of the effective coupling via the ``higher dimensional term". We find that it approaches $1$ when $h \ll M_{\rm pl}/\xi(h \ll M_{\rm pl}/\sqrt{\xi})$, while it goes to zero for $h \gg M_{\rm pl}/\xi (h \gg M_{\rm pl}/\sqrt{\xi})$ for Metric (Palatini) for mulation. 

We can also have other terms for the gauge coupling jump in the presence of the large non-minimal coupling. However, they should behave similarly to have the chiral Standard Model  in the limit $\varphi \to \infty$ (see e.g. \cite{Bezrukov:2009db}), the tree-level gauge coupling should approach a constant so that the derivatives of the induced corrections (e.g. from the QCD potential) to the $\varphi$ potential is vanishing. In the low field range, those ``higher dimensional" terms should behave as the usual higher dimensional terms, and we have the usual perturbative effective theory connecting the Standard Model. 
Similar behavior can be found from threshold effects to the other couplings~\cite{Bezrukov:2014ipa, George:2015nza, Fumagalli:2016lls, Enckell:2016xse, Bezrukov:2017dyv,Shaposhnikov:2020fdv,Poisson:2023tja}.

By including the generic ``higher dimensional terms" in the original Lagrangian, it is well-motivated to consider the scaling of the gluon coupling in the form of 
\beq
\laq{geff}
\boxed{\frac{1}{g^{2}_{\rm eff}}\simeq  
\left\{
\begin{array}{cc}
\frac{1}{g^{2}}+\d g^{-2}&\hspace{-20mm} ~~~~~~~~~~~~~~~~\text{for}~ h\ll M_{\rm pl}/\x \text{~(Metric) or } h\ll M_{\rm pl}/\sqrt{\x} \text{~(Palatini)} \\
\frac{1}{g^{2}}&\hspace{-20mm} \text{for}~ h\gg M_{\rm pl}/\sqrt\x ~\text{(Metric and Palatini)}
\end{array}
\right.}
\eeq
which resembles the behavior $g^{-2} + \d g^{-2} (\partial_\varphi F)^2$. For the Metric formulation, we did include the intermediate range, which is known to be non-perturbative.  We emphasize that if $\d g^{-2}>0$ and not too small, the gluon is stronger coupled at a large $h$ regime. 

In the following we mainly focus on the possibility $\frac{1}{g^{2}}\lesssim 1, \d g^{-2}\gtrsim 1$,   i.e.,
we have a stronger QCD with large $h$ values,and the usual QCD coupling in the small $h$ regime.\footnote{When $h$ is just below the intermediate range, we consider the QCD coupling $g_{\rm eff}\lesssim 0.6$ to connect the Standard Model via the renormalization group running effect. }

\subsection{QCD axion with stronger QCD}

Before we discuss the small instanton in the next section, let us study an application to the QCD axion in the early universe due to the stronger QCD. Although a stronger QCD is not directly detectable in the present Universe, it was known to cause the suppression of the axion abundance if the CP conserving vacuum aligns to the present one~\cite{Dvali:1995ce, Banks:1996ea, Choi:1996fs, Co:2018phi} (see also natural models for this to happen~\cite{Matsui:2020wfx, Kitano:2023mra}), and the suppression of  
the isocurvature perturbation
\cite{Jeong:2013xta}. 
The axion dynamics with a stronger QCD in Higgs inflation by considering a model without a jump of any coupling was also discussed\,\cite{Yin:2022fgo}. In this part, we take into account of the gluon coupling jump to study the dynamics of axion during the early universe, especially during the Higgs inflation. 

To see this, let us introduce the QCD axion $a$ in the form 
\beq
{\cal L}\supset \frac{1}{2}\partial_\m a \partial^\m a  - \frac{a}{16\pi^2 f_a}{\rm tr} G_{\m\n}\tl G^{\m\n}.
\eeq
in the Einstein frame. 
Here $\tl G$ is the dual of the gluon field strength.
We assume for simplicity, $\Omega$ does not appear in the Lagrangian.  
This can be justified at large $h$ values if
we consider that the potential includes 
$V\supset -\lambda \frac{h^2}{\Omega^2} \ab{\F_{\rm PQ}}^2$ in the Jordan frame so that when $h$ is large, the kinematically normalized Peccei-Quinn Higgs, $\F_{\rm PQ}$, gets a $h$-independent negative mass squared and similarly a $h$-independent quartic coupling. The expectation of $\F_{\rm PQ}$ is irrelevant to the $h$ value at the large $h$ limit.\footnote{If this portal coupling is suppressed, we get the vacuum mass $V\supset -\frac{m_{\rm PQ}^2}{\Omega^2} \ab{\F_{\rm PQ}}^2$. This implies that the decay constant gets smaller when the $h$ field is larger. The axion mass becomes even heavier, strengthening the conclusion in this section.} 
At the lower limit the vacuum mass term of $\F_{\rm PQ}$ is assumed to be dominant. In this realization $f_a$ may not be the one in the vacuum. 
We further assume that the symmetric phase of the PQ symmetry is never achieved during the cosmological history so we consider the pre-inflationary PQ breaking scenario. 

During inflation, we have $h\gg M_{\rm pl}/\sqrt{\x}$, and for simplicity we take $h$ as the background field. 
As usual Higgs (or Starobinsky) inflation, at large $h$ values, the theory enters into chiral Standard Model at $h \to \infty$~\cite{Bezrukov:2007ep, Bezrukov:2008ej}, and 
a fermion mass scales with 
\beq
\lim_{h\to \infty} y F\to y \frac{M_{\rm pl}}{\sqrt{\x}}.
\eeq

Then the QCD scale is estimated from the 1loop renormalization group analysis and  is given by 
\beq
\L_{\rm QCD,inf}\sim \frac{M_{\rm pl}}{\sqrt{\x}} e^{-\frac{8\pi^2}{11 g^2}}.
\eeq
Here we estimated the QCD scale from the 1-loop renormalization group running of pure $\rm SU$(3) Yang-Mills from the quark mass scales$\sim M_{\rm pl}/\sqrt\xi$. 
For instance, $\L_{\rm QCD,inf}= \O\(10^{-3}- 10^{-1}\)\frac{M_{\rm pl}}{\sqrt{\x}}$ for $g=1 - 2$ by emphasizing again that $g$ is larger than the IR QCD coupling due to the threshold effect. 
$\(M_{\rm pl}/\sqrt{\xi}\)^4$ is a similar order (neglecting the dimensionless coupling dependence other than $g$ and $\x$) to the inflation potential. 

Assuming that the quark masses are heavier than the QCD scale during inflation we have the axion mass in the form
\beq
m_{a,\rm inf}\sim \frac{\L_{\rm QCD,inf}^2}{f_a}.
\eeq

\paragraph{Suppressing the isocurvature bound}
Axion dark matter faces constraints from isocurvature fluctuations, especially when the axion is light during inflation. If the axion mass is less than the Hubble parameter at that time, it has stochastic fluctuations. These fluctuations spread over scales larger than the horizon and have a scale-invariant power spectrum, causing spatial variations in the axion field values over the horizons.

After inflation ends, the universe continues to expand, and the Hubble parameter decreases. When the Hubble parameter becomes comparable to the axion mass, the axion field starts to oscillate around its potential minimum, accounting for the observed dark matter abundance~\cite{Preskill:1982cy, Dine:1982ah, Abbott:1982af}. However, due to the initial fluctuations during inflation, axion densities vary over large scales even at the time of recombination, leading to isocurvature fluctuations. These fluctuations are strongly constrained by observations of the cosmic microwave background~\cite{Planck:2018jri}.

When $m_{a,\rm inf}\gtrsim H_{\rm inf}$ with $H_{\rm inf}$ being the Hubble parameter during inflation, the axion is stabilized during inflation, and thus the fluctuation is highly suppressed. Namely,
if \beq
m_{a,\rm inf}\gtrsim H_{\rm inf} ~{\rm i.e.}~ e^{-\frac{16\pi^2}{11 g^2}} \gtrsim \frac{f_a}{\sqrt{3}M_{\rm pl}},
\eeq
the isocurvature modes are suppressed and the axion is free from the isocurvature problem. 
For $f_a=10^{15-17}\GEV$ axion, we can satisfy it with $g\gtrsim 1-2$.

\paragraph{QCD axion abundance}

As we will discuss in the next section, depending on the nature of the threshold effect, one  
may or may not change the prediction of the QCD axion abundance. 

With the generic threshold effects, including the ones to the Yukawa couplings, the axion potential minimum during inflation generically shifts to a CP violating one~\cite{Dvali:1995ce, Banks:1996ea, Jeong:2013xta} and thus even if we have axion in the vacuum during inflation, we cannot simply suppress the abundance. The resulting axion abundance depends on the initial misalignment angle, which is threshold effect dependent.

On the other hand, one may have a minimal CP and flavor-violating threshold effects as will discussed in the next section. In this case, one can suppress the abundance as the setup~\cite{ Kitano:2021fdl, Kitano:2023mra}.

\section{Small instanton from non-minimal coupling}
\label{sec: small instanton}
Let us point out with the non-minimal coupling
that a small instanton can be generated even with $\vev{h}= v_{\rm EW}$ at the EW vacuum.

Usually, a stronger QCD with a non-trivial Higgs field value does not mean an enhanced small instanton because in estimating the small instanton we need to integrate out the high momentum modes of the relevant fields, while the zero mode of Higgs, the expectation value, is not large.\footnote{For instance, the small instanton estimation becomes very unclear if we use a gauge kinetic term $\frac{1}{g^2}+\frac{\varphi(x)}{M}$. With high momentum $\varphi$ configuration, the gauge coupling varies everywhere in the real space and the calculation tools are difficult to apply directly. In addition, when $\vev{\varphi_>^2(x)}\gtrsim M^2$, the perpurbative unitarity is violated significantly.}  
In our case, as we will see, the asymptotic behavior of \Eq{geff} smooths 
the short wave modes of the Higgs. 
This will be the key property for our estimation to show the enhanced small instanton.

For simplicity of discussion, we first consider a toy model that has much less problem in the sense of perturbativity.  
Then we come back to the realistic Higgs inflation and discuss the axion physics.

\subsection{Small instanton with gauge coupling jump in a toy model }
\label{Sec: small instanton in toy model}

Let $\varphi$  a real scalar singlet and it does not have a potential, but it couples to gluon via
\beq
\label{eq: L gluon theta}
{\cal L}_{\rm gluon, \theta}=-\frac{1}{2}\(\frac{1}{g^2}+\partial_\varphi F \partial_\varphi F \d g^{-2}\)  {\rm tr} G_{\m\n}G^{\m\n}-\frac{\theta}{32\pi^2}  {\rm tr} G_{\m\n}\tl G^{\m\n}.
\eeq
Here $\theta$ is the strong CP phase. We omit to write down the metric since what we will discuss will be purely field-theoretic. $\partial_\varphi F$ is the one defined in \Eq{Fdef}, and we do not speficy if this is Metric or Palatini formulation for a while. 
This Lagrangian is similar to the one in the Einstein frame of the Higgs non-minimal coupling one with the threshold effect~Eq.\,\eqref{counter}.
This $\f$ does not give mass to any gauge fields, and we do not need to consider the perturbative unitarity from the weak bosons. For simplicity, we also omit gravity and fermions  for a while.

To study the small instanton we need to integrate out the heavy modes, from the momenta much higher than $M_{\rm pl}/\sqrt{\x}$ to the scale of interest, say the EW scale, which is much smaller than $M_{\rm pl}/\sqrt{\x}.$ 
In the intermediated range, the model may become non-perturbative, especially in the case of Metric formulation, and one may consider some UV completion. With or without UV completion, we can always perform the path integral following the Wilsonian approach.

 We will argue that this model has an enhanced small instanton with $ g\gtrsim (g^{-2}+\d g^{-2})^{1/2},$  which comes from the very high momentum modes ($\gg  M_{\rm pl}/\sqrt\x$), although our estimation may not be justified for the intermediate momentum modes. Given the positive path-integral measure of QCD~\cite{Vafa:1984xg}, the contribution from the non-perturbative regime should not cancel the contribution from the high momentum modes.  
 
To study this, let us first consider the Wilsonian effective action in 4D Euclidian space
\beq
e^{i S_{\rm eff}}= 
\int {\cal D}g_> {\cal D}\varphi_>  e^{i S[g_>+g_<,\varphi_>+\varphi_<]}, \laq{WEA}
\eeq 
where 
\beq
S = \int d^4x (\mathcal{L}_{\rm{kin}}[\varphi_>+\varphi_<] + {\cal L}_{\rm gluon, \theta}[g_>+g_<,\varphi_>+\varphi_<]),
\eeq 
with $\mathcal{L}_{\rm{kin}}[\varphi]:=(1/2)(\partial \varphi (x))^2$ and ${\cal L}_{\rm gluon, \theta}$ is given in Eq.~\eqref{eq: L gluon theta}.  The subscripts, $>\AND <$,  denote the high ($[\epsilon\Lambda-\Lambda ]$) and low momentum modes ($[0-\e\Lambda]$). Here $\L$ is the cutoff scale of the action, $S$, with $\e$ being a number smaller than 
$1$. Thus $S_{\rm eff}$ is the effective action with the cutoff scale reduced to $\e \L$. 

To perform the integral of $\e\L-\L$ modes, 
we note that the Higgs field has 
\begin{align}
\label{eq: vev of varphi sq}
    \vev{\varphi_>^2(x)}&:=\frac{\int {\cal D}\varphi_> e^{i\int \cal{L}_{\rm{kin}}(\varphi_>)}\varphi_>(x)^2}{\int {\cal D}\varphi_> e^{i\int \cal{L}_{\rm{kin}}(\varphi_>)}}\\
    &= \int_{\epsilon \Lambda < |k| < \Lambda} {d^4 k\over (2 \pi)^4} {1 \over k^2} \simeq {(1-\epsilon^2)\over (4 \pi)^2} \Lambda^2\sim  {1\over (4 \pi)^2} \Lambda^2
\end{align}
in the path integration over $\varphi_>$, the typical magnitude of the field $\varphi_>$, given by $\varphi_{>,\rm{RMS}}:= \sqrt{\vev{\varphi_>^2(x)}}$, is of the order of $4\pi \Lambda$.

In the following, we consider two cases: $ \L\gg 4\pi M_{\rm pl}/\sqrt{\x}$ and $\L\ll 4\pi M_{\rm pl}/\sqrt \x ( M_{\rm pl}/\x)$ for the Palatini (Metric) formulation. 
\paragraph{When $ \L\gg 4\pi M_{\rm pl}/\sqrt{\x}$,}
 the $\varphi$ dependence in the gauge coupling is highly suppressed, and $\varphi$ is almost a freeparticle justifying the perturbative estimation of Eq.\,\eqref{eq: vev of varphi sq}.
 To verify this rough sketch, we performed a numerical lattice simulation, which is shown in Appendix~\ref{app: numerical lattice simulation}. What we observe in the numerical simulation is indeed that the typical field configuration has large values, $\sim \L$, although it varies rapidly over distances of $1/\L$. 
We remark that if $\L\gg 4\pi M_{\rm pl}/\sqrt{\x}$, {\it the fluctuation for $1/g^2+\d g^{-2} (\partial_\varphi F)^2\simeq 1/g^2$ is neglected} although $\varphi$ varies everywhere. 
The typical coupling of Higgs to the gauge field is exponentially suppressed (see Fig.\,\ref{fig:1}). 
Thus $1/g^2$ is the effective coupling almost everywhere. 
Namely, the threshold effect smooths the Higgs fluctuations in short waves and it decouples in the gauge sector.\footnote{In realistic case, this smoothing is guranteed by the requirement that the Higgs field has a flat potential at the high field value as in the Higgs inflation. }
At a suppressed volume of the field space, $\sim (M_{\rm pl}/\sqrt{\x})/
\sqrt{\varphi^2_{>,\rm RMS}}$, we have the field value below $M_{\rm pl}/\sqrt{\x}$, where our calculation may be invalid. However, the measure in the path integral is suppressed (see Figs.~\ref{fig:3} and \ref{fig:4}). 
This observation allows us to approximate the effective action in the factorized form 
\beq 
 e^{ i S_{\rm eff}}\simeq  \int {\cal D} \varphi_> e^{i{\int d^4x {\cal L}_{\rm kin}}}\times \int {\cal D} g_{>} \left.e^{i {\int d^4x {\cal L}_{\rm gluon,\theta}}}\right|_{\varphi\to \infty}.\laq{path1}
\eeq 
Now $g_>$ integral has been reduced to the simple Yang-Mills path integral with the gauge coupling $g\gtrsim 1$, which is larger than the IR one.

Since the small instanton contribution (c.f.~\cite{Holdom:1982ex, Dine:1986bg, Flynn:1987rs}) comes from the gauge sector, we obtain it by the usual dilute instanton gas approximation~\cite{Callan:1977gz}:
\beq
\laq{vac}
\d_\L V(\theta)
=    \int^{1/\e\L}_{1/\L}{\frac{d \rho' }{\rho'^5} e^{-S_{\rm eff}[1/\rho']}\
  F^{\rm (vac)}[\rho']}
 \cos{\(\theta\)},
\eeq
which gives the potential of the $\theta$ term.  $\rho'$ represents the size modulus of
the instanton solution. 
The dependence of effective action on
$\rho'$ arises from the quantum corrections that are captured by the
gauge coupling constant, $S_{\rm eff}[1/\rho]\approx
2\pi/\a_s[1/\rho]\sim 8\pi^2 /g^2$, which is, importantly, the UV coupling.

Precisely speaking, this gauge coupling is $\L$ dependent.   
Since $\varphi$ with momenta mode much larger than $M_{\rm pl}/\sqrt{\x}$ rarely interacts with the gauge field, 
the gluon gauge coupling runs as usual in the regime. The matching condition is  $g(\mu= \frac{4\pi M_{\rm pl}}{\sqrt{\x}})=g$.
By considering the renormalization group running from the gauge loop, we get the scaling 
\beq e^{-8\pi^2/g^2}\propto \(\frac{4\pi M_{\rm pl}/\sqrt{\x}}{\L}\)^{11}\eeq 
for $\L \gg 4\pi M_{\rm pl}/\sqrt \x.$
Therefore the integral over $\rho'$ is IR dominant. 

\paragraph{When $\L\ll 4\pi M_{\rm pl}/\sqrt \x ( 4\pi M_{\rm pl}/\x)$ for the Palatini (Metric) formulation,} the Higgs field interacts with the gauge field via the higher dimensional term and it decouples. This is another region we can perform the instanton estimation analytically. 
Neglecting the higher dimensional term, we again obtain a factorized form, 
\beq 
 e^{ i S_{\rm eff}}\simeq  \int {\cal D} \varphi_> e^{i{\int d^4x {\cal L}_{\rm kin}}}\times \int {\cal D} g_{>} \left.e^{i {\int d^4x {\cal L}_{\rm gluon,\theta}}}\right|_{\varphi\to 0}.
\eeq 
We can use the dilute gas instanton estimation \Eq{vac}, but with $\a_s =(g^{-2}+\d g^{-2})/4\pi$, which is the one to match the Standard Model prediction. 
This is known to be suppressed, and the we do not consider it. \\

As a consequence, within the calculable regime, the small instanton effect becomes most important at the energy scale of $\L\sim 4\pi M_{\rm pl}/\sqrt{\x},$ corresponding to $\varphi_{>,\rm RMS}$ slightly larger than $M_{\rm pl}/\sqrt{\x}$ at which the gauge coupling jumps. 
 As an order-of-estimate, we can use the value estimated in the UV range to approximate the contribution (or the minimum contribution)
\beq
\laq{dV}
\boxed{V_{\rm inst}(\theta)\sim  10^{-3}\({\frac{8\pi^2}{g^2}}\)^6\frac{M_{\rm pl}^4}{\x^2}e^{-\frac{8\pi^2}{g^2}}\cos[\theta]\times \(\frac{1}{4\pi}\)^{7}}
\eeq
where we used $F[\rho']\approx 10^{-3}\({2\pi / \a_s(1/\rho')}\)^6$ for a single instanton~\cite{tHooft:1976snw}. We also included the factor $\(1/(4\pi))\)^{11-4}$ by taking account of $\rho'^{-1}\sim (4\pi) \x^{-1/2}M_{\rm pl}$ and the renormalization group running.
As a consequence we have an enhanced small instanton effect.

\paragraph{Perturbative unitarity}

In the Palatini formualtion, the perturbative unitarity bound is not very severe even if we introduce gravity coupling. 
 At large field value, $\varphi \gg M_{\rm pl}/\sqrt\x$, $\varphi$ has only the flat potential, and it rarely couples to the gauge field. Indeed, by expanding $\varphi = \varphi_0 +\d \varphi$ the cutoff for the perturbative bound is~\cite{Bauer:2010jg} $\sim e^{ 2\sqrt{\x} |\varphi_0|/M_{\rm pl}} M_{\rm pl}/\sqrt \x,$ which is exponentially larger than the field value of $\varphi_0$.  We may replace $\varphi_0$ by $\varphi_{>,\rm RMS}$ to estimate the perturbative bound, because over the large spatial volume we have the typical value of $\varphi$ to be $\varphi_{>,\rm RMS}$.  
 Thus for the instanton of scale $1/\rho \gg  M_{\rm pl}/\sqrt{\x}$ our discussion so far should work. With $1/\rho \ll  M_{\rm pl}/\sqrt{\x}$, it also works since the cutoff from unitarity is $M_{\rm pl}/\sqrt{\x}$. The vague region is $1/\rho \sim M_{\rm pl}/\sqrt \x$ in which the perturbative unitarity is marginally satisfied or violated, where we perhaps need some UV completion for a more solid calculation. We expect the contribution from this intermediate range does not cancel the contribution that we have studied, since the contribution should have the same sign in the absense of quarks~\cite{Vafa:1984xg}. 
 
 In the Metric formulation, however, by introducing gravity we have a strengthened perturbative bound $\sim \sqrt{6}M_{\rm pl}$ (c.f. Fig.\ref{fig:1} for the condition $\phi \gtrsim M_{\rm pl}/\sqrt{\x} \leftrightarrow \varphi\gtrsim M_{\rm pl}$). This allows a very suppressed range for $\L$ for our estimation to be well justified. Around the range $\L \sim M_{\rm pl}/\sqrt{\x}$ the gauge coupling is already very large and the Higgs field is already fully decoupled from the gauge sector (see Fig.\ref{fig:1}). 

\paragraph{Effects from quarks} By introducing $n_f$ quarks, $q_i$, coupled to the real scalar with $m_i=y_i \phi[\varphi]/\Omega[\varphi]$
the small instanton gets suppressed by the Yukawa couplings. 

We get in the region $\L \gg4\pi M_{\rm pl}/\sqrt{\x}$
\beq 
\d_\L V(\theta)
= 
   \int^{1/\e\L}_{1/\L}{\frac{d \rho' }{\rho'^{5-n_f}}  e^{-S_{\rm eff}[1/\rho']}\
  F^{\rm (vac)}[\rho']} 
 \cos{\(\theta\)} \Pi_{i}\frac{y_i M_{\rm pl}}{\sqrt{\x}}.
\eeq 
Again, the integral dominates at $\L\sim 4\pi M_{\rm pl}/\sqrt{\x}$ and we obtain 
\beq
\d_\L V(\theta)\sim  10^{-3}\({\frac{8\pi^2}{g^2}}\)^6\frac{M_{\rm pl}^4}{\x^2}e^{-\frac{8\pi^2}{g^2}}\cos[\theta] (4\pi)^{-7- n_f/3}\Pi_i {y_i}
\eeq

\subsection{Case of the Higgs inflation and a quality problem for QCD axion}
Now we come back to the Higgs inflation. 
One can estimate the small instanton contribution (apart from the non-perturbative range) 
\beq 
V_{\rm inst}= (2\times 10^8\GEV)^{4} \(\frac{10^{10}}{\x}\)^2\frac{|\det Y_u \det Y_d|}{10^{-18}}g^{-12}e^{-8\pi^2/g^2}\cos[\theta]. \laq{inst2}
\eeq 
We note that $Y_{u,d}$ also jumps~\cite{Bezrukov:2014ipa, George:2015nza, Fumagalli:2016lls, Enckell:2016xse, Bezrukov:2017dyv}. 
The $Y_{u,d}$ jump in general involves the phase shift of the Yukawa coupling. Rotating the phase in the Yukawa coupling introduces the contribution to the strong CP phase via chiral anomaly.\footnote{Given this effect, our discussion by the positive measure~\cite{Vafa:1984xg} to gurantee the absence of the cancellation does not apply. However, since this phase shift effect depends on the threshold effects for the Yukawa couplings, we expect in general the cancellation does not happen.} Therefore, in general, the small 
instanton has a phase misaligned from the IR QCD one unless we consider some specific scenarios. 
 Given those aspects, we consider two possibilities: the general case and specific case.

\paragraph{PQ quality problem with general CP violation}

In this case, the CP-violating small instanton may spoil the Peccei-Quinn solution to the strong CP problem even if we introduce the QCD axion.

Assuming an $\O(1)$ misalignment of the CP phases between the small instanton and conventional IR QCD contributions,
we get the requirement 
\beq
 (2\times 10^8\GEV)^{4} \(\frac{10^{10}}{\x}\)^2\frac{|\det Y_u \det Y_d|}{10^{-18}}g^{-12}e^{-8\pi^2/g^2}\lesssim 10^{-10} \chi  
\eeq 
with $\chi\sim (0.08\GEV)^4$ being the topological susceptibility, and $10^{-10}$ is from the bound from the non-observation of the neutron electric dipole moment ~\cite{Pendlebury:2015lrz}. 
If the heavy degrees of freedom in the PQ sector is much heavier than $M_{\rm pl}/\sqrt{\x}$, this contribution cannot be supprressed by considering accidental PQ symmetry~\cite{DiLuzio:2017tjx, Lee:2018yak, Ardu:2020qmo, Yin:2020dfn}, as discussed in Ref.\,~\cite{Kitano:2021fdl}. 
 For instance, with the parameters that we used, we have $g\lesssim 0.8$. 
On the other hand, if we use 
$
|\det Y_u \det Y_d|=\O(1)
$
by taking account of the parameter jump and assuming the $\O(1)$ coupling in the UV theory, we get $g\lesssim 0.7,$ which is not too different. 

\paragraph{A potential solution to the quality problem and heavy QCD axion}

One possible solution to the quality problem pointed out so far is to assume a minimal flavor and CP violation. 
In other words, we consider $Y_u, Y_d$ to be the complex spurion field for the breaking of the chiral symmetry. 
This property is satisfied at any order of the perturbation either in a small $h$ regime or a large $h$ regime. This is because the ``higher dimensional terms" may also satisfy the symmetry (which is only spontaneously broken by the spurion fields). 
This scenario requires no CP phases for other coupling parameters. For instance, we can include the threshold effect contribution and write the Yukawa interaction in the form $ c(|h|^2) H\bar{Q} Y_u u$.\footnote{We can also consider, e.g., ${\rm det}{(Y_u)} \tl{c}(|h|^2) H\bar{Q} Y_u u$, which satisfies our assumption and is CP violating, in addition to the leading contribution  $ c(|h|^2) H\bar{Q} Y_u u$. The shift to the strong CP phase is expect to be $\O(10^{-10})$ by assuming $\tl{c}$ and $c$ is the same order at $h\to \infty$. Here assume that $c=1,\tl{c}=0$ when $h\to0$. This minimal flavor and CP violating scneairo may be a novel solution to the quality problem. A model building for it may be interesting. } Our assumption says that $c$ does not have CP phase, although $Y_u$ may have. 
This setup still has a strong CP problem because $\arg \det Y_u \det Y_d$ is non-vanishing in general. Thus we will solve it by the QCD axion. 

In this setup interestingly the small instanton must have the CP phase aligned/opposite to the IR one. 
We have a heavy QCD axion (or lighter axion with cancellation between the IR QCD contribution and the small instanton contribution) whose mass can be as high as 
\beq 
m_a\simeq  4\GEV g^{-6}e^{-4\pi^2/g^2}\(\frac{10^{10}}{\x}\)\frac{10^{16}\,\GEV}{f_a}.
\eeq

\paragraph{Perturbative unitarity}
Let us comment on the perturbative unitarity. 
In Higgs inflation we have a more severe perturbative bound compared with the previous toy model. 

Even for the Palatini formulation, we have a bound from 
 the weak boson which has a mass of the form
\beq 
m_V^2\sim g^2 \frac{h^2}{\Omega^2}\to g^2 \frac{M^2_{\rm pl}}{\x}
\eeq
at the large $h$ limit. 
Since Higgs no more interacts with the gauge field, this gives a perturbative unitarity bound of $M_{\rm pl}/\sqrt{\x}$ \cite{Bauer:2010jg}. This coincides with the scale that the gluon coupling becomes strong.

Besides the possible violation of the unitarity bound, we still expect that the small instanton exists with the gauge coupling jump. 
When $h\sim M_{\rm pl}/\sqrt{\x}$ in both formulations, our estimation is marginally justified (see also discussions below \Eq{dV}).
In addition, in Palatini formulation, the violation of the perturbative unitarity is from the weak sector. After integrating out the unitarity-violated weak sector with $\L \gg 4\pi M_{\rm pl}/\sqrt{\x}$, we still have the decopled sectors of gluon and Higgs boson, which are both perturbative.  
A similar small instanton estimation should work. 
As a na\"{i}ve dimensional analysis, we still expect to have the form \eq{inst2} of the small instanton contribution from  the integration in the range $\L\gg 4\pi M_{\rm pl}/\sqrt{\x}$. 

We aslo comment that the perturbative bound from the weak boson is the specific issue for the Standard Model-like Higgs boson (c.f. Sec.\,\ref{sec: conclusions}).

\section{Conclusions}
\label{sec: conclusions}
A scalar field can naturally couple to the Ricci scalar with a non-minimal coupling $\xi \gg 1$ and it is discussed widely in the context of cosmic inflation.
However, the theory is non-renormalizable and the parameters in Standard Model can differ significantly above and below the intermediate scale due to ``higher dimensional terms".
In this paper, we studied the threshold effect on the gauge coupling for the first time relevant to the non-renormalizability and showed that it can induce small instanton and provide stronger QCD in the early Universe. In particular, we found that the dilute instanton gas approximation  can be useful at high energy scales due to the behavior of the threshold effect.
QCD axion quality problem, heavy QCD axion, the suppression of isocurvature perturbation, and the abundance were discussed to be threshold effect dependent. Therefore, a non-minimal coupling may change the axion physics significantly. 
\\

Lastly, let us comment on the generality of our claim by discussing other models with non-minimal couplings. 
In fact, if the PQ Higgs field in the KSVZ model~\cite{Kim:1979if, Shifman:1979if} has a non-minimal coupling, a similar correction to Eq.~\eqref{eq: two loop correction for gluon coupling} arises at the two-loop level due to the presence of the KSVZ quarks (and also the contirbution of the usual Higgs-quark coupling; see below). The same conclusions follow straightforwardly.

More generically, we may consider a scalar field with the non-minimal coupling completely decoupled from any quark in the Jordan frame. 
Even in this case, 
 when we go to the Einstein frame, we have the coupling with the Standard Model sector.  For instance, $\varphi$ couples to the (canonically normalized) Standard Model like Higgs boson since it has a mass term. Other than the derivative couplings to the Higgs boson, from the mass term of the Higgs boson there is a portal coupling $\partial_{\varphi\varphi}\Omega^{-2} m_h^2$. This contributes to the loop diagram for the gluon self-coupling and there is a divergence even in the dimensional regularization and we need a counter term. The thereshold effect naturally exists for the gauge coupling. 
 
Therefore, we consider that the gauge coupling shift occurs more generally, and perhaps once we have a sizable non-minimal coupling for any scalar field we need to be careful of the small instanton and stronger QCD in the early Universe, especially when we study axion.\footnote{The threshold effect also exists for $\SU(2)$ and $\U(1)$ gauge fields in the Standard Model as well as for any other new gauge fields. }


\section*{Acknowledgments}
This work was supported by JSPS KAKENHI Grant No. 20H05851 (W.Y.), 22K14029 (W.Y.), 22H01215 (W.Y.), and 22J21260 (J.W.), and Incentive Research Fund for Young Researchers from Tokyo Metropolitan University (W.Y.).  

\appendix

\section{Numerical lattice simulation of $\varphi$ with high frequency fluctuation}
\label{app: numerical lattice simulation}

In this appendix, we present a simple numerical simulation to demonstrate how the typical field value of $\varphi$ can become sufficiently large, approaching the cut-off scale $\Lambda$ at high energy scales, as discussed in Sec.~\ref{Sec: small instanton in toy model}. We begin by outlining the basic setup and expected results, followed by the presentation of results from the simple lattice simulations.

If $\varphi$ is weakly coupled, we can expand it following the canonical normalization:
\beq
 \varphi (x) = \int \frac{d^3 \vec{k}}{\(2\pi\)^3\sqrt{2k}}  \hat a_{\vec{k}}  e^{i \vec k \vec x}+h.c.
\eeq
 $\hat{a}_{\vec k}$ is the annihilation operator for momentum $\vec{k}$ mode. 
The Minkovski fluctuation can be derived as 
\beq
\laq{proba}
\vev{\varphi_{\vec{k}} \varphi_{-\vec{k}}}=\vev{\ab{\varphi_{\vec{k}}}^2}= \frac{1}{2 k} V.
\eeq
Here $\varphi_{\vec{k}}$ is the Fourier transformation of $\varphi(x)$.
We used $(2\pi)^3\delta^3(\vec{k}-\vec{k})= V.$
To see the field configuration of $\varphi$ from $\e\L$ to $\L$, we may generate the values of $\varphi_{\vec{k}}$ with $\e\L<k<\L$ at random following the probability distribution of \Eq{proba}. This procedure is usually used in classical Lattice simulation to take into account the quantum fluctuation~\cite{Felder:2000hq, Figueroa:2020rrl, Figueroa:2021yhd}.\footnote{We also emphasize the cosmological simulation, the super horizon fluctuation may be different. The difference of the initial conditions would cause significantly different results~\cite{Gonzalez:2022mcx, Kitajima:2023kzu}.}

\begin{figure}[t!]
\begin{center}  
\includegraphics[width=105mm]{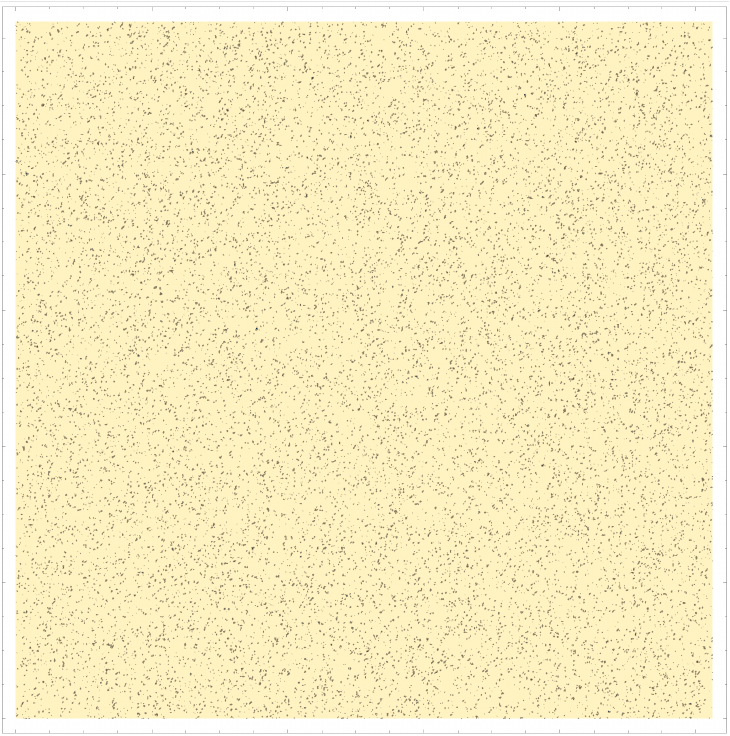}
\end{center}
\caption{2D slice of the configuration of the Higgs field with $\L L/(2\pi)=256 $ on $1024^3$ lattices.  We generate the modes in the range $\L/\e<k<\L$. The black dots denote the region with $\phi< \L/(30\times 2\pi)$, which is highly subdominant. If $ \L/(30\times 2\pi) \sim M_{\rm pl}/\sqrt{\x}$, in the yellow region the gauge coupling is constant and is $g$, while black dots have different gauge coupling.
The perturbative unitarity is violated in black dots, while it is well preserved in the yellow region. Our instanton estimation is valid in the yellow regimes, while the contribution from black dots is neglected. }
\label{fig:3}
\end{figure}
\begin{figure}[t!]
\begin{center}  
\includegraphics[width=105mm]{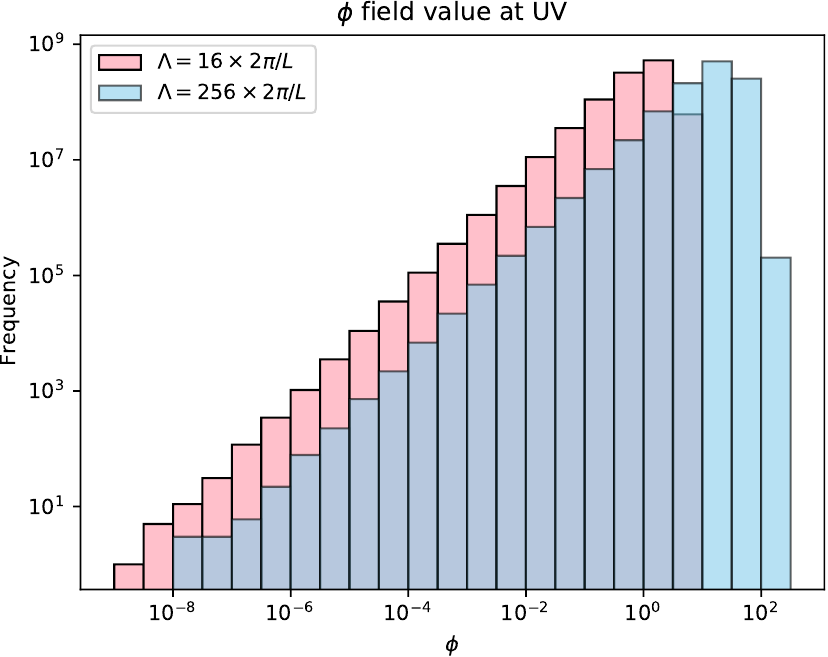}
\end{center}
\caption{Histogram of the $\varphi$ field values in the simulation (see also Fig.\,\ref{fig:3}). We use $\L=256 (16) \times 2\pi/L$ for the blue (pink) one.}
\label{fig:4}
\end{figure}

The numerical lattice simulation is performed to show this behavior in Fig.~\ref{fig:3}. We also show the histogram of the $\varphi$ filed values in Fig~\ref{fig:4}.

\bibliography{ref}

\end{document}